\pdfoutput=1
\documentclass[11pt]{article}
\usepackage[]{acl}
\usepackage{times}
\usepackage{enumitem}
\usepackage{tikz}
\usetikzlibrary{shapes, arrows.meta, positioning}
\usepackage{booktabs}
\usepackage{multirow}
\usepackage{latexsym}
\usepackage{amsmath}  
\usepackage{amssymb}   
\usepackage{tikz} 
\usepackage{hyperref}
\usepackage{xcolor}
\usepackage[T1]{fontenc}
\usepackage[utf8]{inputenc}
\usepackage{microtype}
\usepackage{amsmath,amssymb}           
\usepackage{algpseudocode}         
\usepackage{inconsolata}
\usepackage{graphicx}
\usepackage{subcaption, algorithm2e, url, float, etoolbox, adjustbox, pgf,booktabs,verbatim}
\usepackage{xcolor,graphicx,colortbl}

\title{Bridging Attribution and Open-Set Detection using Graph-Augmented Instance Learning in Synthetic Speech}

\author{
 \textbf{Mohd Mujtaba Akhtar\textsuperscript{1}\thanks{Equal contribution as a first author.}},
 \textbf{Girish\textsuperscript{2}\footnotemark[1]},
 \textbf{Farhan Sheth\textsuperscript{3}\footnotemark[1]} and
 \textbf{Muskaan Singh\textsuperscript{4}\thanks{Corresponding: \href{mailto:m.singh@ulster.uk.in}{m.singh@ulster.uk.in}}} \\
 \textsuperscript{1}Veer Bahadur Singh Purvanchal University, India \\
 \textsuperscript{2}UPES, India \\
 \textsuperscript{3}Manipal University Jaipur, India \\
 \textsuperscript{4}Ulster University, UK
}

\begin{document}
\maketitle

\begin{abstract}
We propose a unified framework for not only attributing synthetic speech to its source but also for detecting speech generated by synthesizers that were not encountered during training. This requires methods that move beyond simple detection to support both detailed forensic analysis and open-set generalization. To address this, we introduce \textbf{\texttt{SIGNAL}}, a hybrid framework that combines speech foundation models (SFMs) with graph-based modeling and open-set-aware inference. Our framework integrates Graph Neural Networks (GNNs) and a k-Nearest Neighbor (KNN) classifier, allowing it to capture meaningful relationships between utterances and recognize speech that doesn't belong to any known generator. It constructs a query-conditioned graph over generator class prototypes, enabling the GNN to reason over relationships among candidate generators, while the KNN branch supports open-set detection via confidence-based thresholding. We evaluate \textbf{\texttt{SIGNAL}} using the DiffSSD dataset, which offers a diverse mix of real speech and synthetic audio from both open-source and commercial diffusion-based TTS systems. To further assess generalization, we also test on the SingFake benchmark. Our results show that \textbf{\texttt{SIGNAL}} consistently improves performance across both tasks, with Mamba-based embeddings delivering especially strong results. To the best of our knowledge, this is the first study to unify graph-based learning and open-set detection for tracing synthetic speech back to its origin.

\end{abstract}

\section{Introduction \& Background}
Synthetic Speech Detection (SSD) plays a critical role in safeguarding digital communication, enabling systems to identify and mitigate the risks posed by highly realistic, machine-generated voices \cite{todisco2019asvspoof,wu2015asvspoof}. With the advent of advanced text-to-speech (TTS) and voice conversion (VC) models, synthetic speech has reached a level of fidelity that closely mimics natural human prosody and timbre \cite{ren2020fastspeech,kong2020diffwave}. While such advancements drive progress in accessibility and personalization \cite{cooper2020zero}, they also introduce new vulnerabilities in the form of audio-based impersonation, fraud, and misinformation \cite{yi2023audio}. Consequently, the ability to detect and analyze synthetic speech is vital not just for security but also for preserving trust in human-AI interaction. The past few years have witnessed remarkable advancements in neural speech synthesis, driven by diffusion-based models, expressive TTS systems, and multilingual voice conversion techniques. State-of-the-art models such as VALL-E \cite{ju2024naturalspeech}, NaturalSpeech 3 \cite{ju2024naturalspeech}, and Voicebox \cite{le2023voicebox} have demonstrated the ability to generate speech that not only mimics speaker identity but also captures fine-grained acoustic attributes such as emotion, prosody, and expressiveness. These models, often built upon large-scale speech-language pretraining, leverage powerful architectural backbones including transformers \cite{li2019neural}, state-space models \cite{gu2023mamba}, and denoising diffusion processes \cite{ren2020fastspeech}. Their increasing accessibility through open-source implementations and commercial APIs has made synthetic speech generation more ubiquitous than ever. However, this rapid progress also underscores the growing complexity of the detection task, particularly in settings where the generation model is unknown or unseen at inference time. \par
Despite the increasing attention on synthetic speech detection, most existing methods frame SSD as a binary classification problem, focusing solely on distinguishing real from synthetic audio rather than identifying the generative source \cite{shin2024hm,huang2024self}. However, most existing methods fail to generalize when faced with previously unseen TTS or voice-conversion models, exhibiting significant performance degradation under open-set conditions \cite{guo2024audio,stan2025tada}. This limitation constrains their practical utility in real-world forensic scenarios, where source attribution and robust detection of out-of-distribution speech are critical. While recent work has advanced synthetic speech detection, a key gap remains: the ability to both identify the source TTS model and detect speech from unseen generators. Source attribution is increasingly important in forensic and regulatory settings, where knowing the origin of synthetic audio matters. At the same time, real-world systems must handle open-set scenarios, where new or unknown models may appear at test time. Bridging these two challenges is essential for building more reliable and generalizable detection frameworks. 
\noindent As the core focus of our study in DiffSSD, we explore a range of speech foundation models (SFMs), and \textit{hypothesize that combining relational modeling via Graph Neural Networks (GNNs)—where a query-conditioned graph is formed over generator class prototypes (nodes) and refined through attention-based message passing—with open-set inference through k-Nearest Neighbors (KNN) offers an effective strategy for jointly tackling source attribution and unseen generator detection.} This approach is motivated by the complementary strengths of the two components: GNNs model class-level interactions among embedding nodes, capturing subtle relational cues for generator discrimination, while the KNN branch provides a lightweight yet robust mechanism for handling open-set scenarios via confidence-based thresholding. To validate this hypothesis, we perform extensive evaluations using diverse SFM embeddings—including Whisper, UniSpeech, ECAPA, and Mamba—under both seen and unseen generator conditions. Drawing inspiration from prior work on explainable source attribution \cite{mishra2025towards}, the use of graph-based structures for deepfake detection \cite{febrinanto2025vision}, and Graph Attention Networks for spoofing detection \cite{tan2025dual}, To adress this, we introduce \textbf{\texttt{SIGNAL:}} \textbf{S}peech \textbf{I}nference via \textbf{G}raph \textbf{N}etworks and \textbf{A}ugmented \textbf{L}earning — a unified framework that performs GNN-based node classification alongside post-hoc KNN filtering to address both tasks simultaneously. To the best of our knowledge, this is the first work to explore a hybrid GNN–KNN approach for source tracing and open-set detection in synthetic speech.

\noindent\textbf{Our key contributions are as follows:}
\vspace{-0.1cm}
\begin{itemize}
\item We propose \textbf{\texttt{SIGNAL}}, a novel hybrid framework combining Graph Neural Networks (GNNs) for relational modeling with k-Nearest Neighbors (KNN) for open-set inference, tailored for joint source attribution and unseen generator detection in synthetic speech.
\vspace{-0.5cm}
\item We perform a large-scale benchmarking of diverse Speech Foundation Models (SFMs)—including Whisper, UniSpeech, ECAPA, and Mamba—under both seen and unseen generator settings, revealing their complementary behavior across tasks.
\vspace{-1mm}
\item To the best of our knowledge, this is the first study to investigate graph-based posthoc reasoning on SFM embeddings for open-set synthetic speech forensics, establishing strong performance on both DiffSSD and SingFake benchmarks.
\end{itemize}
\noindent \textit{Resources for this study are available at: {\url{https://github.com/Helixometry/SIGNAL.git}}}

\begin{figure*}[!h]
    \centering
    \includegraphics[width=0.972\textwidth]{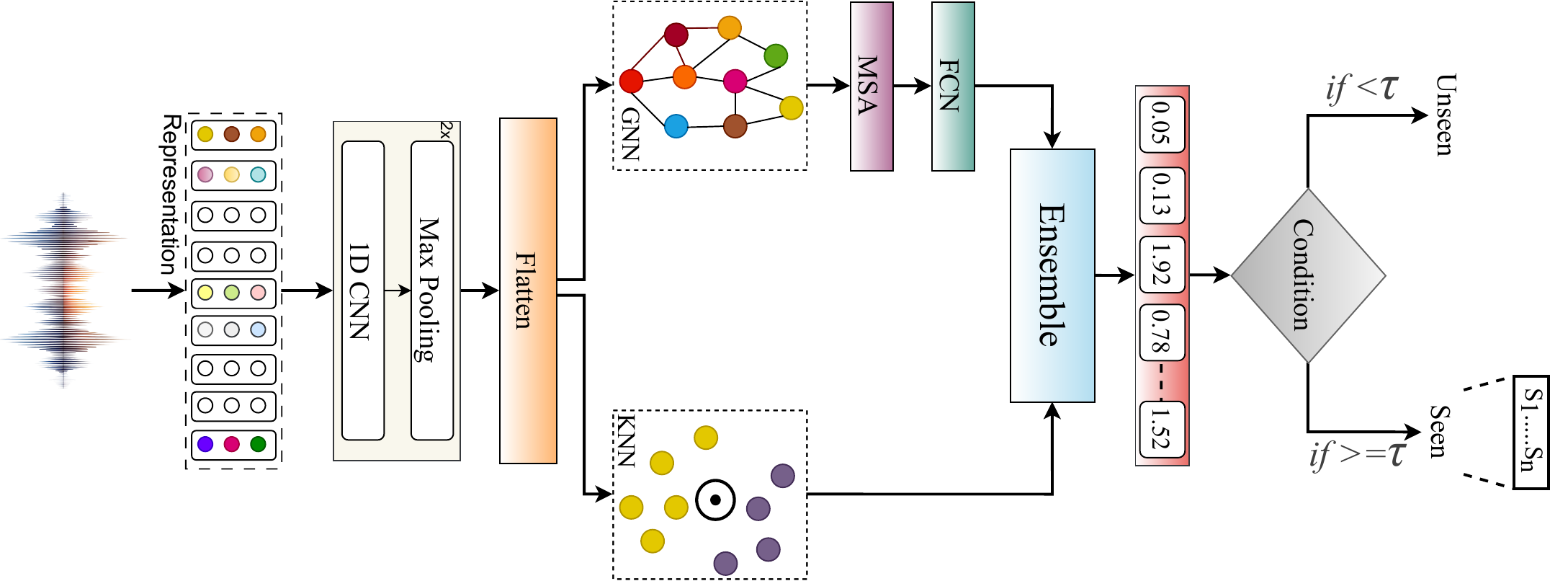}
    \caption{Proposed framework: \textbf{\texttt{SIGNAL}}. The model extracts representations, followed by parallel reasoning via a Graph Attention Network (GAT) and a K-Nearest Neighbors (KNN) module. The outputs are fused via an ensemble head. The final routing decision is based on a confidence threshold ($\tau = 0.5$), directing samples to either seen ($S_1 \ldots S_n$) or unseen class predictions.}
    \label{fig-1}
\end{figure*}

\section{Representations}
In this section, we detail the speech foundation models (SFMs) used to extract utterance-level embeddings for downstream processing.

\noindent We utilize x-vector\footnote{\url{https://huggingface.co/speechbrain/spkrec-xvect-voxceleb}} \cite{8461375} and ECAPA-TDNN\footnote{\url{https://huggingface.co/speechbrain/spkrec-ecapa-voxceleb}} \cite{desplanques20_interspeech} as speaker recognition models. Both are based on time-delay neural networks and are trained on the VoxCeleb1+2 datasets. ECAPA extends the standard x-vector architecture with Res2Net modules and SE blocks, yielding improved performance over x-vector, which itself significantly surpasses traditional i-vector baselines. x-vector contains approximately 4.2M parameters. For monolingual PTMs, we employ WavLM\footnote{\url{https://huggingface.co/microsoft/wavlm-base}} \cite{chen2022wavlm}, UniSpeech-SAT\footnote{\url{https://huggingface.co/microsoft/unispeech-sat-base}} \cite{chen2022unispeech}, and wav2vec 2.0\footnote{\url{https://huggingface.co/facebook/wav2vec2-base}} \cite{baevski2020wav2vec}, all in their base versions trained on 960 hours of English audio from LibriSpeech. WavLM and UniSpeech-SAT are top-performing PTMs on the SUPERB benchmark, with WavLM trained using masked speech modeling and denoising objectives, while UniSpeech-SAT adopts a multi-task framework incorporating speaker-aware contrastive learning. wav2vec 2.0 is optimized via a contrastive loss to distinguish true from distractor codebook entries. All three models have similar parameter sizes: WavLM (94.7M), UniSpeech-SAT (94.68M), and wav2vec 2.0 (95.04M). We further include Whisper\footnote{\url{https://huggingface.co/openai/whisper-base}} \cite{radford2023robust}, a multilingual ASR model trained on 680K hours of diverse audio. Whisper uses a transformer-decoder architecture and performs robustly across languages and tasks in a zero-shot setting. We use the base version with 74M parameters and extract representations from the encoder layer. In addition, we consider Audio-MAMBA\footnote{\url{https://github.com/SarthakYadav/audio-mamba-official}} \cite{yadav24_interspeech}, a state space model trained on AudioSet to reconstruct masked spectrogram patches. Its SSM backbone captures both temporal dependencies and spectral detail effectively. We use its tiny (4.8M), small (17.9M), and base (69.3M) variants to study scalability effects. \par
\noindent Resampling to 16 KHz is done for the audio samples before passing it to the FMs. We obtain fixed-length embeddings by applying average pooling to the final hidden state of each frozen model. The resulting embedding dimensions are: 512 for Whisper and x-vector; 768 for WavLM, UniSpeech-SAT, and wav2vec 2.0; and 3840 for Audio-MAMBA (base version). ECAPA-TDNN produces 192-dimensional embeddings.

\section{Modeling Pipeline}

In this section, we detail the modeling pipeline for individual representations and the proposed hybrid framework for source tracing and unseen generator detection. We use Fully Connected Networks (FCN) and Convolutional Neural Networks (CNN) as downstream models for individual representation-based modeling. Further, we propose \textbf{\texttt{SIGNAL}}, which integrates Graph Neural Networks (GNNs) with a K-Nearest Neighbors (KNN) module, enabling it to leverage both global relational structure and local instance-level similarity. Figure-\ref{fig-1} illustrates the overall architecture.

\begin{table*}[!h]
\setlength{\tabcolsep}{2pt}
\centering
\scriptsize
\begin{tabular}{l|ccc|ccc|ccc|ccc|ccc|ccc}
\toprule
\textbf{} & \multicolumn{9}{c|}{\textbf{FCN}} & \multicolumn{9}{c}{\textbf{CNN}} \\
\cmidrule(lr){2-10} \cmidrule(lr){11-19}
\multicolumn{1}{c|}{\multirow{3}{*}{\textbf{PTMs}}} &
\multicolumn{3}{c|}{\textbf{DEV}} &
\multicolumn{3}{c|}{\textbf{TEST}} &
\multicolumn{3}{c|}{\textbf{ID/OOD}} &
\multicolumn{3}{c|}{\textbf{DEV}} &
\multicolumn{3}{c|}{\textbf{TEST}} &
\multicolumn{3}{c}{\textbf{ID/OOD}} \\
\cmidrule(lr){2-4} \cmidrule(lr){5-7} \cmidrule(lr){8-10}
\cmidrule(lr){11-13} \cmidrule(lr){14-16} \cmidrule(lr){17-19}
\multicolumn{1}{c|}{} &
ACC \(\uparrow\) & F1 \(\uparrow\) & EER \(\downarrow\) & ACC \(\uparrow\) & F1 \(\uparrow\) & EER \(\downarrow\) & ACC \(\uparrow\) & F1 \(\uparrow\) & EER \(\downarrow\) &
ACC \(\uparrow\) & F1 \(\uparrow\) & EER \(\downarrow\) & ACC \(\uparrow\) & F1 \(\uparrow\) & EER \(\downarrow\) & ACC \(\uparrow\) & F1 \(\uparrow\) & EER \(\downarrow\)  \\
\midrule
\multicolumn{19}{c}{\textbf{Baseline}} \\
\midrule
Whisper        & \underline{80.75} & \underline{79.3} & \underline{7.78}  & 74.52 & 72.66 & \textbf{9.77}  & 52.35 & 51.17 & 41.09 & \textbf{83.10} & \textbf{81.40} & \underline{5.71}  & \underline{79.45} & \underline{78.13} & \underline{7.81}  & 54.12 & 52.85 & 30.11 \\
Unispeech      & 70.58 & 69.12 & 27.20 & 72.00 & 70.26 & 29.70 & 37.54 & 36.77 & 53.99 & 78.35 & 76.86 & 20.34 & 68.96 & 67.35 & 23.45 & 38.81 & 38.05 & 40.32 \\
x-vector       & 66.91 & 65.42 & 19.86 & 69.47 & 67.52 & 22.67 & 43.83 & 42.80 & 52.20 & 74.15 & 72.90 & 16.42 & 67.82 & 65.58 & 21.39 & 45.50 & 44.39 & 37.78 \\
wav2vec 2.0       & 76.82 & 75.45 & 8.73  & 69.46 & 67.31 & 12.88 & 41.90 & 41.08 & 44.33 & 78.13 & 76.42 & 6.49  & 79.19 & 77.44 & 9.40  & 43.43 & 42.62 & 32.66 \\
wavLM          & 78.24 & 76.81 & 12.48 & 73.16 & 71.07 & 15.53 & 44.74 & 43.65 & 52.49 & 80.60 & 78.25 & 9.06  & 76.44 & 74.21 & 11.37 & 46.38 & 45.34 & 39.11 \\
ECAPA          & 63.77 & 62.55 & 17.28 & 62.31 & 60.09 & 25.29 & 32.33 & 31.63 & 57.80 & 71.92 & 69.87 & 13.13 & 64.15 & 63.29 & 20.13 & 33.56 & 32.84 & 41.45 \\
MAMBA-T        & 69.35 & 68.12 & 15.87 & \underline{76.01} & \textbf{75.40} & 19.08 & \underline{59.23} & \underline{58.13} & 46.59 & 80.15 & 78.17 & 11.37 & 72.23 & 71.04 & 11.79 & \underline{60.94} & \underline{59.92} & 35.00 \\
MAMBA-S        & 72.01 & 70.66 & 10.86 & 71.04 & 69.16 & 13.19 & 57.33 & 56.21 & \underline{40.56} & 81.22 & 79.56 & 7.82  & 74.51 & 73.18 & 9.80  & 59.32 & 58.11 & \underline{29.64} \\
MAMBA-B        & \textbf{81.68} & \textbf{80.29} & \textbf{7.63}  & \textbf{76.10} & \underline{74.35} & \underline{10.74} & \textbf{62.91} & \textbf{61.50} & \textbf{36.14} & \underline{82.90} & \underline{80.11} & \textbf{5.59}  & \textbf{80.08} & \textbf{78.47} & \textbf{7.78}  & \textbf{64.62} & \textbf{63.35} & \textbf{26.27} \\
\bottomrule
\end{tabular}
\caption{Performance metrics (ACC, F1, EER) across different PTMs under FCN and CNN backbones. Top-performing scores are in \textbf{bold}, runner-up scores are \textit{underlined}. All tables in the study follow the same formatting.}
\label{tab-1}
\end{table*}

\subsection{Individual Representation Modeling}

To establish strong baselines, we first evaluate standard classifiers with individual pre-trained speech representations, we use FCN and CNN backbones. The CNN consists of two 1D convolutional layers with 64 and 128 filters (kernel size = 3), followed by ReLU activation and max-pooling (pool size = 2). The output is flattened and passed through a dense layer of 128 neurons, followed by a softmax output layer. The FCN model uses the same dense block as the CNN, excluding convolutional layers.

\section*{Proposed framework : \textbf{\texttt{SIGNAL}}}

We propose, \textbf{\texttt{SIGNAL}} for the joint task of source attribution and unseen generator detection in synthetic speech. The architecture of \textbf{\texttt{SIGNAL}} is shown in Figure~\ref{fig-1}.

\noindent \textbf{Representation Encoding:} Let $\mathbf{x} \in \mathbb{R}^{T \times F}$ denote the input audio signal, where $T$ is the number of frames and $F$ is the feature dimension. This signal is passed through a frozen Speech Foundation Model (SFM), yielding a fixed-length utterance embedding:

\begin{equation*}
\mathbf{z}_0 = \mathrm{SFM}(\mathbf{x}) \in \mathbb{R}^{d_0}
\end{equation*}

\noindent We then project $\mathbf{z}_0$ into a lower-dimensional latent space using a CNN encoder $f_{\text{cnn}}$:

\begin{equation*}
\mathbf{z} = f_{\text{cnn}}(\mathbf{z}_0) \in \mathbb{R}^{d}, \quad \text{where } d = 64
\end{equation*}

\noindent The encoder $f_{\text{cnn}}$ comprises two 1D convolutional layers with ReLU activations and max-pooling, followed by a dense projection layer.

\noindent \textbf{GNN Head:} To capture class-level relationships and enhance source attribution, we design a graph-based attention module centered around the input query and known generator classes. The graph consists of:
\begin{itemize}
    \item A \textbf{query node} representing the encoded utterance embedding $\mathbf{z}$.
    \item \textbf{Class nodes} represented by $N$ learnable prototype vectors $\{\mathbf{e}_1, \dots, \mathbf{e}_N\}$, each corresponding to a seen TTS generator.
    \item \textbf{Edges} connecting the query to each class node, allowing the model to assess similarity and perform reasoning over them.
\end{itemize}
\noindent \textbf{Why a GNN for attribution:} We adopt a GNN for attribution because source prediction benefits from reasoning over relationships among generator classes, not just per-class similarity scores. By propagating information across class nodes, the GNN captures relative structure that is difficult to represent with independent classifiers.

\noindent The query embedding $\mathbf{z}$ is first projected into a latent space using a learnable transformation:
\begin{equation*}
\mathbf{s} = \mathbf{W}_s \mathbf{z}, \quad \mathbf{s} \in \mathbb{R}^d
\end{equation*}
where $\mathbf{W}_s \in \mathbb{R}^{d \times d}$ is a trainable weight matrix.

\noindent This projected vector $\mathbf{s}$ is then \textit{added to each class prototype} to produce query-aware node features:
\begin{equation*}
\tilde{\mathbf{e}}_i = \mathbf{e}_i + \mathbf{s}, \quad \forall i \in \{1, \dots, N\}
\end{equation*}

\noindent These $N$ modified node embeddings are passed through a \textbf{multi-head self-attention} mechanism, where each node attends to the others and refines its own representation:
\[
\tilde{\mathbf{e}}_i' = \mathrm{MultiHeadAttn}(\tilde{\mathbf{e}}_i, \{\tilde{\mathbf{e}}_j\}_{j=1}^{N})
\]

\noindent To compute class scores, each updated node is projected to a scalar logit:
\begin{equation*}
\ell_i = \mathbf{w}^\top \tilde{\mathbf{e}}_i', \quad \forall i \in \{1, \dots, N\}
\end{equation*}

\noindent The final class probabilities are obtained via a softmax over the logits:
\begin{equation*}
\mathbf{p}_{\text{GNN}} = \mathrm{softmax}([\ell_1, \dots, \ell_N])
\end{equation*}

\noindent To estimate the model’s uncertainty in attribution, we compute the attention entropy over the predicted distribution:
\begin{equation*}
\mathcal{H}_{\text{attn}} = -\sum_{i=1}^{N} p_{\text{GNN},i} \log p_{\text{GNN},i}
\label{eq:attn_entropy}
\end{equation*}

\noindent Intuitively, low entropy implies confident attribution to a specific generator, while high entropy suggests the model is uncertain and attention was spread more uniformly across classes.

\noindent \textbf{KNN Branch:} To handle open-set scenarios, we introduce a KNN module that performs instance-based reasoning in the embedding space. After training $f_{\text{cnn}}$, we collect all training embeddings $\{\mathbf{z}_j\}$ and fit a distance-weighted KNN classifier.

\noindent At test time, for a query $\mathbf{z}$, the KNN output is computed as:

\begin{equation*}
\mathbf{p}_{\text{KNN}} = \frac{\sum_{k=1}^{K} w_k \cdot \mathbf{y}_k}{\sum_{k=1}^{K} w_k}, \quad w_k = \frac{1}{\|\mathbf{z} - \mathbf{z}_k\|_2^2 + \epsilon}
\end{equation*}

\noindent where $\mathbf{z}_k$ are the $K$ nearest neighbors with corresponding labels $\mathbf{y}_k$, and $\epsilon$ is a small constant to ensure numerical stability.

\noindent \textbf{Ensemble Fusion and Open-Set Detection:} We fuse the GNN and KNN predictions via convex combination:

\begin{equation*}
\mathbf{p}_{\text{ens}} = \alpha \cdot \mathbf{p}_{\text{GNN}} + (1 - \alpha) \cdot \mathbf{p}_{\text{KNN}}, \quad \alpha \in [0, 1]
\end{equation*}

\noindent To determine whether a sample belongs to a known or unknown generator, we use confidence-based routing with threshold $\tau$ and optionally an entropy-based uncertainty signal:
\begin{itemize}
    \item \textbf{Confidence thresholding (main):} if $\max(\mathbf{p}_{\text{ens}}) < \tau$, the sample is labeled as \textit{unseen}.
    \item \textbf{Entropy thresholding (optional):} if $\mathcal{H}_{\text{attn}} > \tau_e$, the sample is labeled as \textit{unseen}.
\end{itemize}

\noindent Unless stated otherwise, we use confidence thresholding as our default decision rule and ablate $\tau$ in Sec.~\ref{ablatinghgho} (Fig.~\ref{fig:enter-label}); the entropy criterion is included as an auxiliary uncertainty signal and not used for primary model selection.

\noindent \textbf{Training Objective:} The GNN head is trained using cross-entropy loss over the seen classes:

\begin{equation*}
\mathcal{L}_{\text{CE}} = -\sum_{i=1}^{N} y_i \log \left( \mathbf{p}_{\text{GNN},i} \right)
\end{equation*}

\noindent The KNN module is non-parametric and requires no training.

\begin{table*}[!hbt]
\setlength{\tabcolsep}{10pt}
\centering
\scriptsize
\begin{tabular}{l|ccc|ccc|ccc}
\toprule
\multirow{3}{*}{\textbf{PTM}} & \multicolumn{3}{c}{\textbf{DEV}} & \multicolumn{3}{c}{\textbf{TEST}} & \multicolumn{3}{c}{\textbf{ID/OOD}} \\
\cmidrule(lr){2-4} \cmidrule(lr){5-7} \cmidrule(lr){8-10}
& \textbf{ACC} $\uparrow$ & \textbf{F1} $\uparrow$ & \textbf{EER} $\downarrow$
& \textbf{ACC} $\uparrow$ & \textbf{F1} $\uparrow$ & \textbf{EER} $\downarrow$
& \textbf{ACC} $\uparrow$ & \textbf{F1} $\uparrow$ & \textbf{EER} $\downarrow$ \\

\midrule
\multicolumn{10}{c}{\textbf{KNN}} \\
\midrule
Whisper   & \cellcolor{blue!27}\underline{92.59} & \cellcolor{blue!27}\underline{90.15} & \cellcolor{blue!27}\underline{5.52}  & \cellcolor{blue!10}85.41 & \cellcolor{blue!10}84.50 & \cellcolor{blue!27}\underline{7.83}  & 60.36 & 58.15 & \cellcolor{blue!10}25.56 \\
Unispeech & 86.14 & 85.06 & 16.05 & 75.13 & 73.32 & 19.99 & 43.14 & 41.37 & 30.14 \\
X-vector  & 82.63 & 80.27 & 14.32 & 76.39 & 74.54 & 19.47 & 49.22 & 47.61 & 32.91 \\
wav2vec 2.0  & 88.41 & 86.38 & 7.18  & \cellcolor{blue!27}\underline{87.76} & \cellcolor{blue!27}\underline{85.91} & \cellcolor{blue!10}8.22  & 50.37 & 48.04 & 27.33 \\
WavLM     & 90.43 & 88.26 & 8.94  & 82.43 & 80.83 & 10.52 & 49.19 & 47.21 & 30.36 \\
ECAPA     & 78.37 & 76.15 & 12.39 & 72.88 & 70.43 & 18.19 & 38.01 & 36.95 & 33.57 \\
Mamba-T   & 89.32 & 87.19 & 11.05 & 80.91 & 79.30 & 10.45 & \cellcolor{blue!10}67.91 & \cellcolor{blue!10}65.36 & 28.39 \\
Mamba-S   & \cellcolor{blue!10}91.26 & \cellcolor{blue!10}89.07 & \cellcolor{blue!10}7.11  & 80.66 & 78.81 & 8.44  & \cellcolor{blue!27}\underline{69.95} & \cellcolor{blue!27}\underline{68.09} & \cellcolor{blue!27}\underline{23.68} \\
Mamba-B   & \cellcolor{blue!45}\textbf{94.05} & \cellcolor{blue!45}\textbf{93.28} & \cellcolor{blue!45}\textbf{5.32}  & \cellcolor{blue!45}\textbf{89.41} & \cellcolor{blue!45}\textbf{87.19} & \cellcolor{blue!45}\textbf{7.62}  & \cellcolor{blue!45}\textbf{72.56} & \cellcolor{blue!45}\textbf{70.24} & \cellcolor{blue!45}\textbf{22.83} \\
\midrule
\multicolumn{10}{c}{\textbf{GNN}} \\
\midrule
Whisper   & \cellcolor{blue!45}\textbf{97.29} & \cellcolor{blue!27}\underline{96.11} & \cellcolor{blue!45}\textbf{4.37}  & \cellcolor{blue!45}\textbf{93.64} & \cellcolor{blue!27}\underline{93.21} & \cellcolor{blue!27}\underline{5.78}  & 64.13 & 62.88 & \cellcolor{blue!10}22.46 \\
Unispeech & 91.26 & 89.26 & 15.89 & 81.38 & 75.18 & 17.54 & 47.28 & 45.86 & 28.85 \\
X-vector  & 86.95 & 82.94 & 12.33 & 80.23 & 79.25 & 17.48 & 52.34 & 50.58 & 27.61 \\
wav2vec 2.0  & 92.05 & 85.20 & 5.74  & \cellcolor{blue!27}\underline{93.40} & \cellcolor{blue!10}90.79 & \cellcolor{blue!10}6.92  & 54.01 & 52.06 & 24.38 \\
WavLM     & 93.62 & 91.67 & 6.86  & \cellcolor{blue!10}89.37 & 86.33 & 8.25  & 53.27 & 51.52 & 28.46 \\
ECAPA     & 82.36 & 78.25 & 10.39 & 75.03 & 72.71 & 17.09 & 42.69 & 40.54 & 30.57 \\
Mamba-T   & 94.29 & 91.22 & 8.96  & 85.51 & 82.64 & 9.23  & \cellcolor{blue!27}\underline{71.92} & \cellcolor{blue!10}69.66 & 26.29 \\
Mamba-S   & \cellcolor{blue!27}\underline{96.41} & \cellcolor{blue!10}92.64 & \cellcolor{blue!27}\underline{5.18}  & 87.26 & 85.42 & 7.34  & \cellcolor{blue!45}\textbf{72.75} & \cellcolor{blue!27}\underline{70.12} & \cellcolor{blue!27}\underline{21.41} \\
Mamba-B   & \cellcolor{blue!10}96.35 & \cellcolor{blue!45}\textbf{96.15} & \cellcolor{blue!10}5.29  & \cellcolor{blue!27}\underline{93.40} & \cellcolor{blue!45}\textbf{93.32} & \cellcolor{blue!45}\textbf{5.62}  & \cellcolor{blue!10}71.22 & \cellcolor{blue!45}\textbf{71.02} & \cellcolor{blue!45}\textbf{19.28} \\
\midrule
\multicolumn{10}{c}{\textbf{KNN + GNN}} \\
\midrule
Whisper   & \cellcolor{blue!27}\underline{97.56} & \cellcolor{blue!27}\underline{96.29} & \cellcolor{blue!27}\underline{3.38}  & \cellcolor{blue!10}94.19 & \cellcolor{blue!10}92.22 & \cellcolor{blue!27}\underline{4.33}  & 76.67 & 74.14 & \cellcolor{blue!27}\underline{17.10} \\
Unispeech & 92.31 & 91.25 & 9.03  & 82.63 & 80.36 & 13.72 & 63.15 & 61.90 & 22.91 \\
X-vector  & 88.19 & 86.99 & 7.79  & 81.59 & 80.41 & 13.20 & 64.66 & 62.35 & 20.96 \\
Wav2vec 2.0  & 93.46 & 90.34 & 3.95  & \cellcolor{blue!27}\underline{94.38} & \cellcolor{blue!27}\underline{92.37} & \cellcolor{blue!10}5.39  & 72.05 & 70.26 & 19.65 \\
WavLM     & 95.09 & 93.34 & 5.15  & 91.21 & 89.84 & 6.71  & 74.28 & 71.47 & 21.35 \\
ECAPA     & 83.27 & 81.06 & 6.55  & 77.61 & 75.29 & 15.07 & 62.36 & 60.32 & 23.91 \\
Mamba-T   & 95.64 & 93.12 & 4.16  & 87.37 & 85.28 & 7.94  & \cellcolor{blue!10}80.17 & \cellcolor{blue!10}78.66 & 20.31 \\
Mamba-S   & \cellcolor{blue!10}96.10 & \cellcolor{blue!10}94.37 & \cellcolor{blue!10}3.91  & 89.24 & 87.32 & 5.90  & \cellcolor{blue!27}\underline{86.54} & \cellcolor{blue!27}\underline{84.12} & \cellcolor{blue!10}17.36 \\
Mamba-B   & \cellcolor{blue!45}\textbf{98.11} & \cellcolor{blue!45}\textbf{96.86} & \cellcolor{blue!45}\textbf{2.33}  & \cellcolor{blue!45}\textbf{95.52} & \cellcolor{blue!45}\textbf{94.21} & \cellcolor{blue!45}\textbf{4.32} & \cellcolor{blue!45}\textbf{88.91} & \cellcolor{blue!45}\textbf{86.53} & \cellcolor{blue!45}\textbf{14.78} \\
\bottomrule
\end{tabular}
\caption{Performance comparison across different Pretrained Models (PTMs) using GNN, KNN, and their combination. The \textcolor{blue}{Blue} gradient indicates performance from highest to lowest. Table~\ref{tab-4} uses the same scheme.}
\label{tab-2}
\end{table*}

\section{Experiments}
\label{Exp-result}
\subsection{Benchmark Dataset}
\label{corpus}

We conduct our study on the Diffusion-Based Synthetic Speech Dataset (DiffSSD) 
\cite{bhagtani2025diffssd}, which contains high-quality synthetic and real speech samples. This dataset is specifically designed to evaluate models in both source tracing and unseen generator detection tasks for synthetic speech. In total, DiffSSD includes around 200 hours of labeled audio, with 70,000 synthetic and 24,226 real speech samples. The real speech in DiffSSD comes from LibriSpeech, which contributes 11,126 samples, and LJ Speech, with 13,100 samples. The synthetic portion is produced by ten TTS systems—eight open-source generators (GradTTS, OpenVoiceV2, ProDiff, WaveGrad2, Xttsv2, YourTTS, DiffGAN-TTS, and UnitSpeech) and two commercial tools (ElevenLabs and PlayHT). Each synthetic sample is created using a set of 5,000 English text lines covering everyday topics such as conversations, weather, quotes, and general descriptions. These text lines were generated using ChatGPT-3.5 and were carefully filtered to avoid repetition. The dataset is divided into three parts: the training set with 31,690 samples, the validation set with 7,423 samples, and the test set with 54,613 samples. This split supports two types of evaluation—closed-set, where all generator types are included during training, and open-set, where the test set includes synthetic speech from generators not seen during training, such as the commercial tools PlayHT and ElevenLabs.
\newline
\noindent \textbf{Training details:}
We follow the predefined train/dev/test splits provided by DiffSSD. Models are trained on the training split and selected using the dev split; all final numbers are reported on the held-out test split (including the ID/OOD setting). We train for up to 50 epochs using Adam with a learning rate of $1\times 10^{-3}$ and batch size 32, and apply early stopping based on dev performance.

\subsection{Evaluation Metrics}
\vspace{-0.34cm}
To evaluate the performance of our proposed framework, we employ widely adopted metrics: Accuracy (ACC), F1-score (F1), and Equal Error Rate (EER). Accuracy provides a straightforward measure of correct classifications. F1-score, as the harmonic mean of precision and recall, offers a balanced view of performance in class-imbalanced scenarios. EER, represents the point at which false acceptance and false rejection rates are equal, making it particularly insightful for open-set evaluation. These metrics have been used in \citet{chetia2025towards} on deepfake detection and source attribution, ensuring methodological consistency across domains and reinforcing the generalizability of our evaluation framework.

\subsection{Experimental Results}
Table~\ref{tab-1} presents performance using FCN and CNN classifiers on individual pre-trained representations. Among the baselines, Mamba-B consistently outperforms other PTMs across all splits. Under the CNN setup, it achieves ACC: 83.27\%, F1: 82.63\%, and EER: 4.26\% on the DEV set, while maintaining reasonable generalization to in-domain/out-of-domain (ID/OOD) settings (ACC: 69.01\%, EER: 17.97\%). Other representations like Whisper and wav2vec 2.0 show competitive performance in seen scenarios but degrade substantially in the presence of unseen generators. ECAPA and UniSpeech yield relatively lower results, reflecting the limitations of speaker-centric and monolingual embeddings in generalizing to diverse generative artifacts. These findings indicate that while CNN-based classifiers can leverage strong pre-trained representations for attribution, they are still limited in open-set scenarios. \newline
\noindent\textbf{Baseline clarification:} We evaluate three categories of methods: (i) Attribution-only models, which assume all test samples originate from known generators and output a closed-set class prediction (e.g., GNN-only); (ii) Open-set-only models, which focus on detecting unseen generators without fine-grained attribution among seen classes (e.g., KNN-only); and
(iii) \emph{Unified} models, which jointly perform attribution and open-set detection. \textbf{\texttt{SIGNAL}} belongs to the third category and explicitly combines GNN-based attribution with KNN-based open-set reasoning. Table~\ref{tab-2} reports all three settings, enabling direct comparison of the trade-offs between attribution accuracy, open-set detection, and their joint optimization. The standalone KNN branch improves over CNN/FCN baselines on ID/OOD detection—for example, Mamba-B achieves an ID/OOD EER of 22.83\%, compared to 26.27\% from CNN. The GNN head further enhances attribution performance by modeling class-level relations, and improves F1 and EER for several PTMs including Whisper and Wav2Vec 2.0. Notably, Whisper’s EER drops from 5.52\% (KNN) to 4.37\% (GNN) on the DEV set. The combined GNN+KNN configuration—our proposed \textbf{\texttt{SIGNAL}} framework—achieves the best performance across all settings. With Mamba-B, \textbf{\texttt{SIGNAL}} attains ACC: 98.11\%, F1: 96.86\%, and EER: 2.33\% on the DEV set, while demonstrating strong generalization to unseen generators with ID/OOD ACC: 88.91\% and EER: 14.78\%. Even smaller models like Mamba-T show noticeable gains under this hybrid setup, validating the complementary strengths of graph reasoning and local neighborhood-based inference. \par
\begin{figure}[!h]
    \centering
    \subfloat[]{%
        \includegraphics[width=0.22\textwidth]{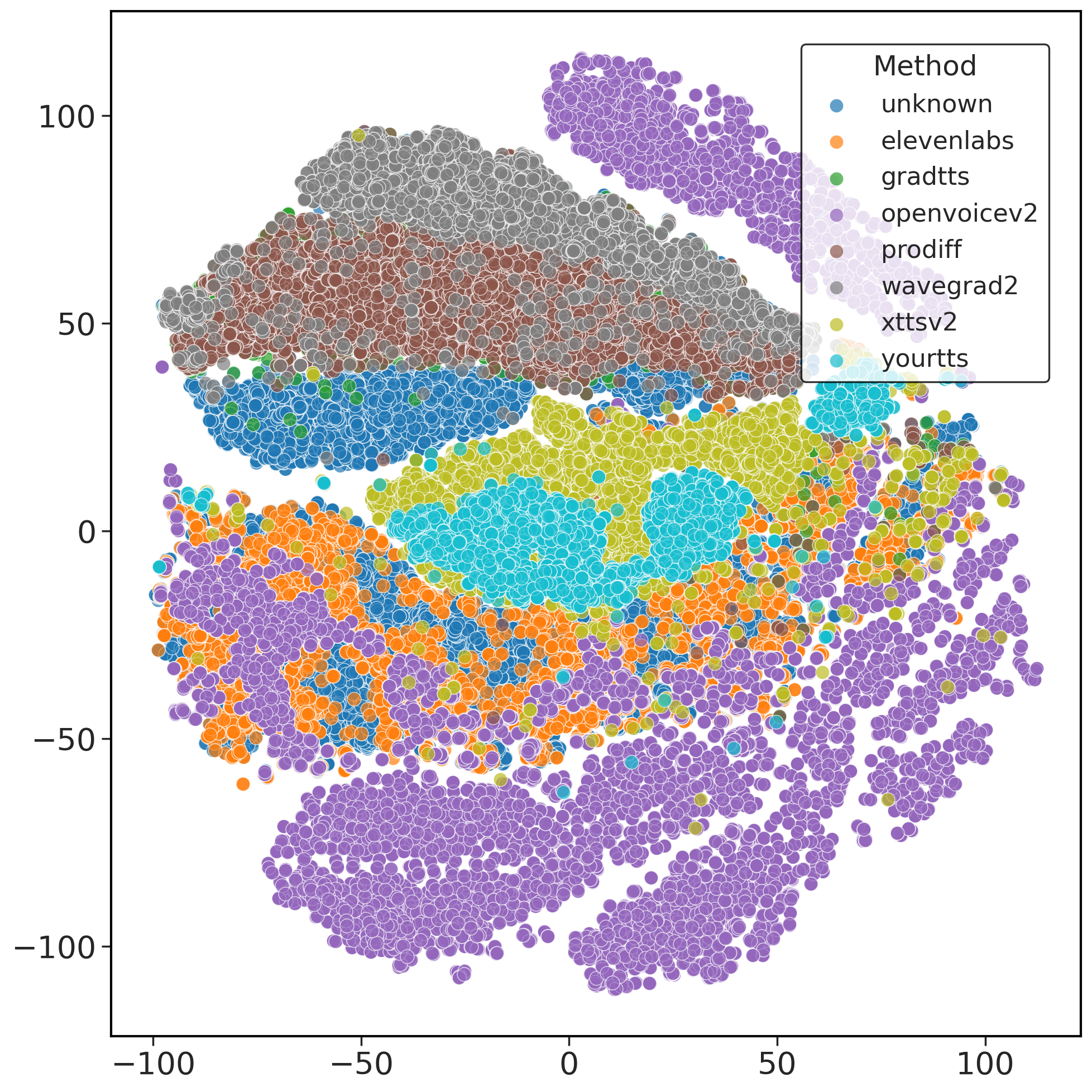}
        \label{fig:tsne_a}
    }
    \hspace{0.3mm}
    \subfloat[]{%
        \includegraphics[width=0.22\textwidth]{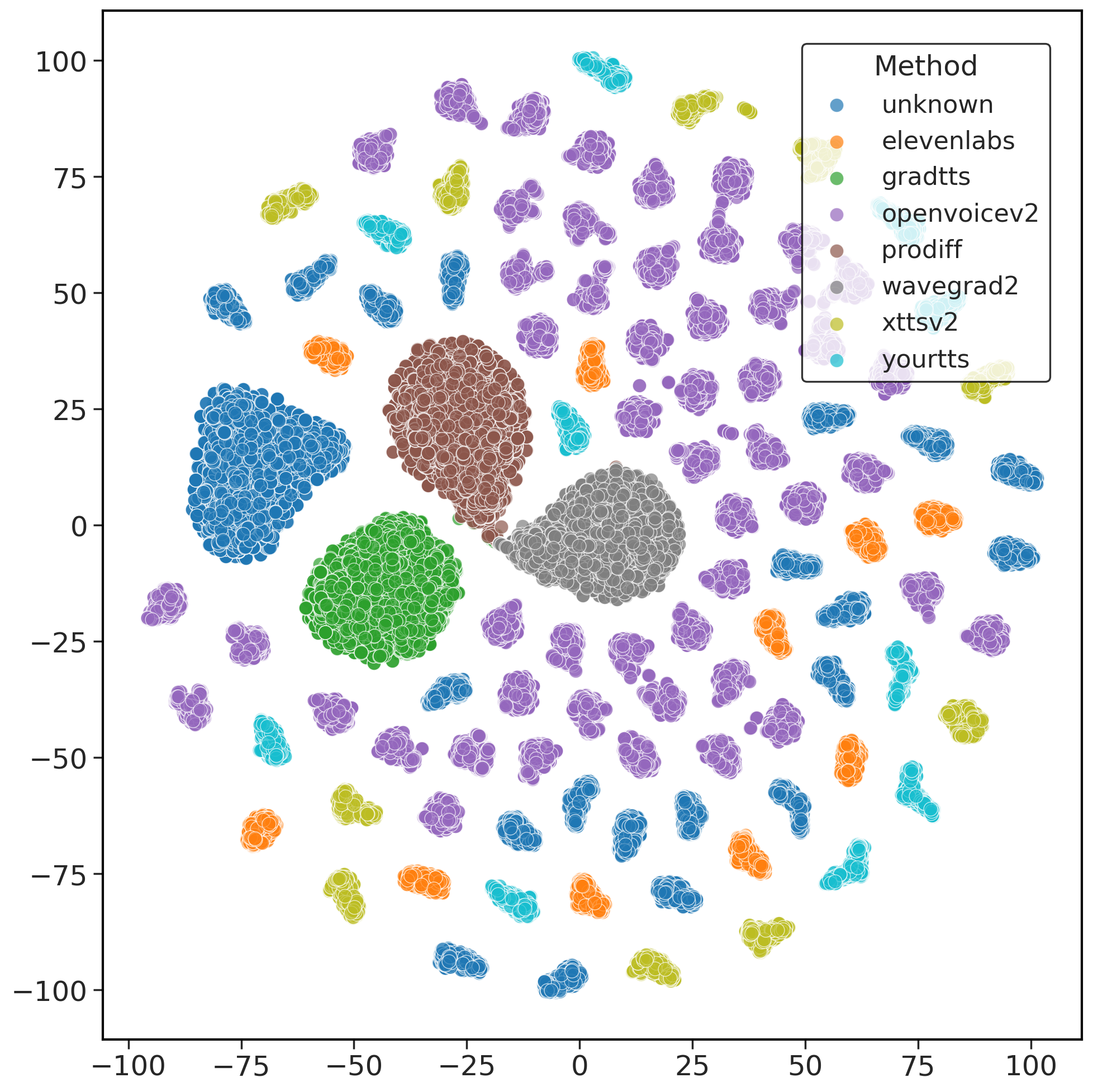}
        \label{fig:tsne_b}
    }
     \hspace{0.3mm}
    \subfloat[]{%
        \includegraphics[width=0.22\textwidth]{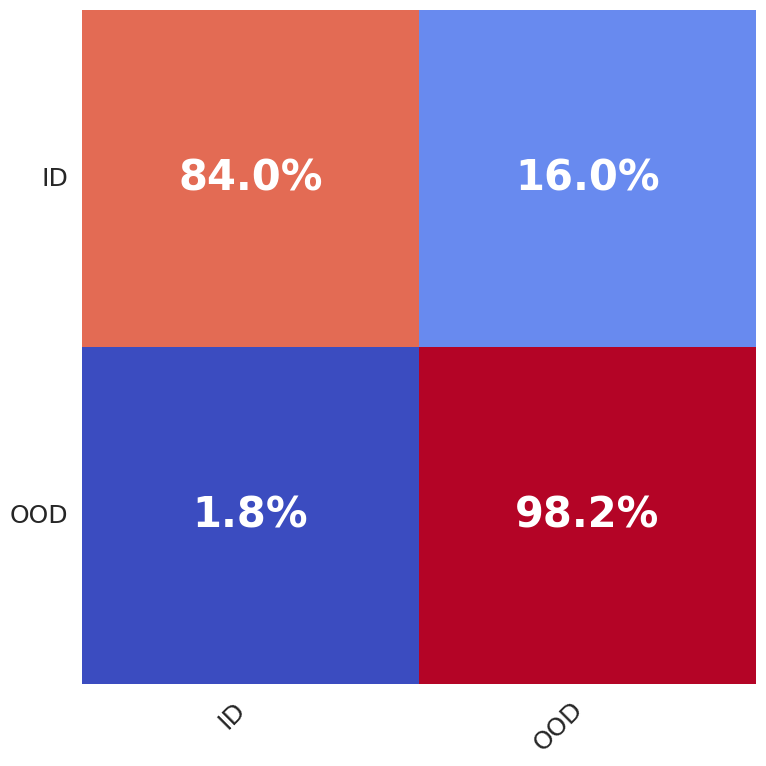}
        \label{fig:tsne_c}
    }
     \hspace{0.3mm}
    \subfloat[]{%
        \includegraphics[width=0.22\textwidth]{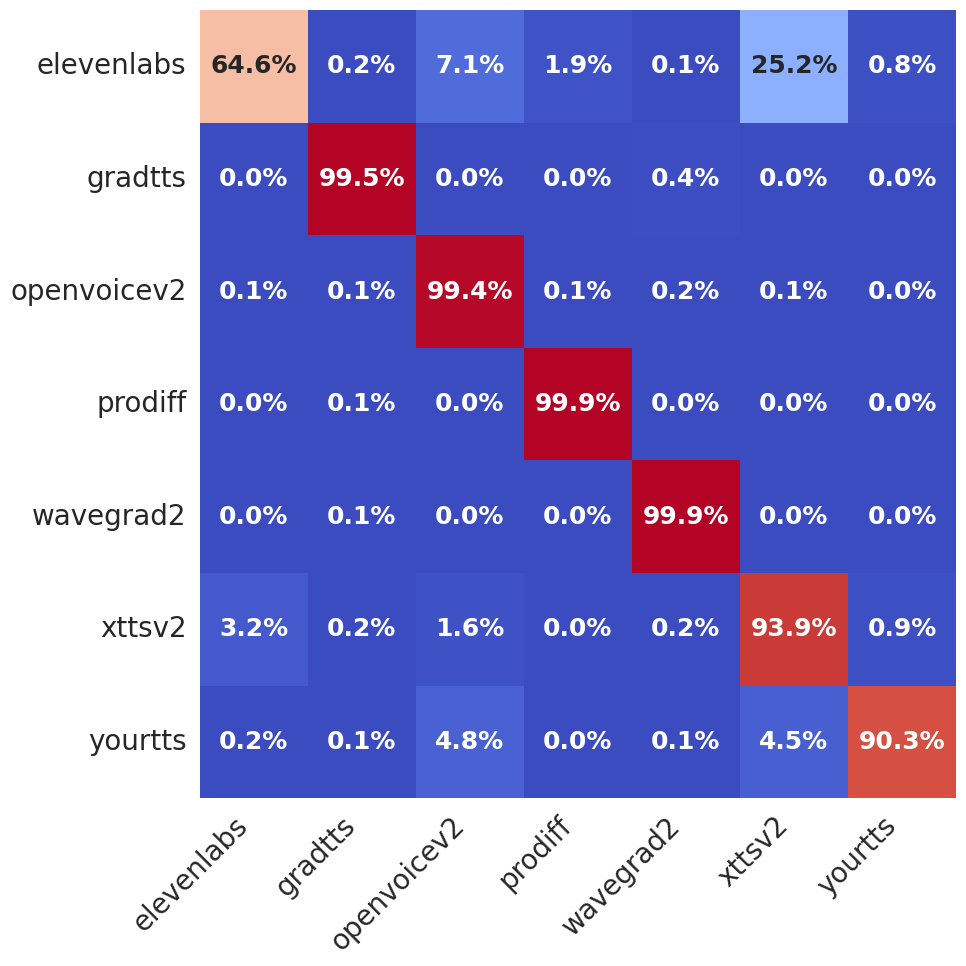}
        \label{fig:tsne_d}
    }
    % \caption{t-SNE Plots- (a) MAMBA-B (test diffssd) (b) MAMBA-B (GNN)  (c) MAMBA-B ID/ODD GNN+KNN (Diffssd) (d) MAMBA-B GNN+KNN test diffssd}
   \caption{Subfigure a and b depict t-SNE : (a) shows the raw embedding space on the DiffSSD test set, while (b) illustrates enhanced class separation after GNN-based refinement. While subfigure c and d present confusion matrices (c) shows ID/OOD separation on DiffSSD using the GNN+KNN ensemble, and (d) shows full test set attribution performance.}

    \label{fig:2}
\end{figure}

\begin{table}[!hbt]
\setlength{\tabcolsep}{5pt}
\centering
\scriptsize
\begin{tabular}{l|ccc|ccc}
\toprule
\multirow{2}{*}{\textbf{PTM}} & \multicolumn{3}{c|}{\textbf{DEV}} & \multicolumn{3}{c}{\textbf{ID/OOD}} \\
\cmidrule(lr){2-4} \cmidrule(lr){5-7}
& ACC & F1 & EER & ACC & F1 & EER \\
\midrule
\multicolumn{7}{c}{\textbf{FCN}} \\
\midrule
Whisper    & 69.62 & 67.42 & 10.64 & 58.16 & 56.67 & 26.79 \\
Unispeech  & 48.59 & 47.11 & 20.35 & 56.31 & 54.08 & 33.89 \\
x-vector   & 71.12 & 69.73 & 12.16 & \underline{64.26} & \underline{62.36} & 32.61 \\
wav2vec 2.0   & 66.25 & 64.56 & 11.28 & 57.45 & 54.34 & 28.50 \\
wavLM      & 42.81 & 41.19 & 23.30 & 59.25 & 57.13 & 34.11 \\
ECAPA      & 70.85 & 68.34 & 13.98 & 62.58 & 60.24 & 36.22 \\
MAMBA-T    & 70.92 & 69.46 & 9.04  & 60.62 & 59.37 & 30.52 \\
MAMBA-S    & \underline{79.56} & \underline{78.02} & \underline{6.22}  & 63.76 & 61.52 & \underline{25.22} \\
MAMBA-B    & \textbf{80.39} & \textbf{79.14} & \textbf{5.49}  & \textbf{65.77} & \textbf{63.28} & \textbf{23.14} \\
\midrule
\multicolumn{7}{c}{\textbf{CNN}} \\
\midrule
Whisper    & 71.64 & 68.91 & 8.66  & 61.03 & 60.17 & \underline{21.26} \\
Unispeech  & 50.05 & 49.88 & 18.05 & 59.16 & 58.28 & 27.91 \\
x-vector   & 73.04 & 71.33 & 10.23 & \underline{67.39} & \underline{66.58} & 25.80 \\
wav2vec 2.0   & 67.71 & 66.17 & 9.38  & 60.74 & 59.95 & 24.52 \\
wavLM      & 45.67 & 44.05 & 20.67 & 62.65 & 62.02 & 26.13 \\
ECAPA      & 72.34 & 70.12 & 11.28 & 65.89 & 64.90 & 29.89 \\
MAMBA-T    & 72.55 & 70.99 & 8.29  & 63.74 & 62.97 & 25.12 \\
MAMBA-S    & \underline{81.64} & \underline{80.16} & \underline{5.01}  & 67.05 & 66.12 & 21.34 \\
MAMBA-B    & \textbf{83.27} & \textbf{82.63} & \textbf{4.26}  & \textbf{69.01} & \textbf{68.12} & \textbf{17.97} \\
\bottomrule
\end{tabular}
\caption{Performance of various Pretrained Models (PTMs) on the SingFake dataset using FCN and CNN classifiers. Models are trained on DiffSSD and directly evaluated on SingFake (zero-shot setting).}
\label{tab-3}
\end{table}

\begin{table*}[!hbt]
\setlength{\tabcolsep}{2pt}
\centering
\scriptsize
\begin{tabular}{l|ccc|ccc|ccc|ccc|ccc|ccc}
  \toprule
  \multicolumn{1}{c|}{\multirow{2}{*}{\textbf{PTM}}}
    & \multicolumn{3}{c}{\textbf{DEV}}
    & \multicolumn{3}{c}{\textbf{ID/OOD}}
    & \multicolumn{3}{c}{\textbf{DEV}}
    & \multicolumn{3}{c}{\textbf{ID/OOD}}
    & \multicolumn{3}{c}{\textbf{DEV}}
    & \multicolumn{3}{c}{\textbf{ID/OOD}} \\
  \cmidrule(lr){2-4}
  \cmidrule(lr){5-7}
  \cmidrule(lr){8-10}
  \cmidrule(lr){11-13}
  \cmidrule(lr){14-16}
  \cmidrule(lr){17-19}
  \multicolumn{1}{c|}{}
    & ACC $\uparrow$   & F1 $\uparrow$    & EER $\downarrow$   & ACC $\uparrow$  & F1 $\uparrow$    & EER $\downarrow$  
    & ACC $\uparrow$   & F1 $\uparrow$    & EER $\downarrow$  & ACC $\uparrow$  & F1 $\uparrow$    & EER $\downarrow$  
    & ACC $\uparrow$   & F1 $\uparrow$    & EER $\downarrow$   & ACC $\uparrow$   & F1 $\uparrow$    & EER $\downarrow$ \\
  \midrule
  \multicolumn{7}{c}{\textbf{KNN}}
    & \multicolumn{6}{c}{\textbf{GNN}}
    & \multicolumn{6}{c}{\textbf{KNN + GNN}} \\
  \cmidrule(lr){1-7}
  \cmidrule(lr){8-13}
  \cmidrule(lr){14-19}
  Whisper   & 75.51 & 74.31 & \cellcolor{blue!10}7.57   & 62.97 & 61.04 & 18.26
             & 77.45 & 76.44 & \cellcolor{blue!10}6.17  & 64.98 & 64.07 & 16.19
             & 85.52 & 84.41 & \cellcolor{blue!10}4.92  & 70.46 & 69.61 & 14.13 \\
  Unispeech & 70.27 & 69.76 & 17.71 & 61.19 & 60.26 & 19.17
             & 72.42 & 71.34 & 15.48 & 63.01 & 62.26 & 17.08
             & 80.38 & 79.17 & 14.73 & 68.32 & 67.36 & 15.67 \\
  X-vector  & 76.39 & \cellcolor{blue!10}75.76 & 9.54  & 67.31 & \cellcolor{blue!10}66.34 & 16.35
             & \cellcolor{blue!10}78.85 & \cellcolor{blue!10}77.91 & 8.73  & \cellcolor{blue!27}\underline{71.77} & \cellcolor{blue!10}70.26 & 15.36
             & \cellcolor{blue!10}87.34 & \cellcolor{blue!10}86.29 & 7.18  & \cellcolor{blue!27}\underline{77.64} & \cellcolor{blue!27}\underline{76.86} & 14.88 \\
  wav2vec 2.0  & 71.77 & 70.13 & 9.01  & 61.72 & 60.19 & 15.61
             & 73.45 & 71.97 & 8.31  & 64.65 & 63.81 & 14.34
             & 80.89 & 79.76 & 6.64  & 70.06 & 69.15 & \cellcolor{blue!10}13.83 \\
  wavLM     & 74.57 & 73.01 & 17.83 & 63.44 & 61.52 & 18.87
             & 76.54 & 75.07 & 15.66 & 66.79 & 66.12 & 16.27
             & 85.17 & 84.32 & 14.39 & 72.41 & 71.33 & 16.02 \\
  ECAPA     & 73.89 & 72.11 & 10.32 & \cellcolor{blue!27}\underline{68.28} & 65.22 & 21.36
             & 78.22 & 77.18 & 9.74  & 70.05 & 69.27 & 19.56
             & 86.61 & 85.48 & 8.75  & 76.04 & 75.20 & 19.07 \\
  MAMBA-T   & \cellcolor{blue!10}76.65 & 75.32 & 7.99  & 64.15 & 62.78 & \cellcolor{blue!10}15.25
             & 78.22 & 77.44 & 7.38  & 67.97 & 67.15 & \cellcolor{blue!10}14.28
             & 87.14 & 86.19 & 6.76  & 73.96 & 72.29 & 13.92 \\
  MAMBA-S   & \cellcolor{blue!27}\underline{84.11} & \cellcolor{blue!27}\underline{83.28} & \cellcolor{blue!27}\underline{4.87}  & \cellcolor{blue!10}67.64 & \cellcolor{blue!27}\underline{66.91} & \cellcolor{blue!27}\underline{13.74}
             & \cellcolor{blue!27}\underline{86.12} & \cellcolor{blue!27}\underline{85.16} & \cellcolor{blue!27}\underline{4.58}  & \cellcolor{blue!10}71.42 & \cellcolor{blue!27}\underline{70.42} & \cellcolor{blue!27}\underline{11.96}
             & \cellcolor{blue!27}\underline{89.29} & \cellcolor{blue!27}\underline{87.95} & \cellcolor{blue!27}\underline{4.29}  & \cellcolor{blue!10}77.26 & \cellcolor{blue!10}76.49 & \cellcolor{blue!27}\underline{10.38} \\
  MAMBA-B   & \cellcolor{blue!45}\textbf{86.79} & \cellcolor{blue!45}\textbf{84.78} & \cellcolor{blue!45}\textbf{4.19}  & \cellcolor{blue!45}\textbf{70.14} & \cellcolor{blue!45}\textbf{69.07} & \cellcolor{blue!45}\textbf{10.71}
             & \cellcolor{blue!45}\textbf{87.20} & \cellcolor{blue!45}\textbf{86.04} & \cellcolor{blue!45}\textbf{4.18}  & \cellcolor{blue!45}\textbf{73.47} & \cellcolor{blue!45}\textbf{72.52} & \cellcolor{blue!45}\textbf{9.52}
             & \cellcolor{blue!45}\textbf{92.28} & \cellcolor{blue!45}\textbf{90.99} & \cellcolor{blue!45}\textbf{3.81}  & \cellcolor{blue!45}\textbf{79.66} & \cellcolor{blue!45}\textbf{78.55} & \cellcolor{blue!45}\textbf{7.92} \\
  \bottomrule
\end{tabular}
\caption{Performance of different PTMs on the SingFake dataset using KNN, GNN, and the proposed GNN+KNN hybrid \textbf{\texttt{SIGNAL}}.}
\label{tab-4}
\end{table*}
\noindent These outcomes reaffirm that success in synthetic speech detection depends on more than just model size or architecture—it requires the right pairing of representations and reasoning. \textbf{\texttt{SIGNAL}} reflects this principle by effectively combining structured graph learning and local similarity-based inference. Furthermore, Figure~\ref{fig:2} provides a comprehensive visual analysis of \textbf{\texttt{SIGNAL}} behavior. Subfigures~\ref{fig:tsne_a} and~\ref{fig:tsne_b} present t-SNE projections of the learned embedding space. In~\ref{fig:tsne_a}, the raw DiffSSD test embeddings show overlapping regions across generator classes, indicating limited separability. After GNN-based refinement in~\ref{fig:tsne_b}, the clusters become more compact and well-separated, highlighting the GNN’s effectiveness in modeling inter-class relationships for better attribution. Subfigures~\ref{fig:tsne_c} and~\ref{fig:tsne_d} show confusion matrices under ID/OOD and full test settings, respectively. \newline
\noindent\textbf{Additional Experiments:} To evaluate the generalization capability of our framework beyond synthetic speech, we experiment on SingFake \cite{zang2024singfake}, a benchmark dataset for singing voice deepfake detection (SVDD). The corpus comprises 28.93 hours of bonafide and 29.40 hours of deepfake clips across five languages and 40 singers, including diverse generative models and musical contexts. We evaluate \textbf{\texttt{SIGNAL}} in a zero-shot setting—models trained on DiffSSD are directly tested on SingFake without any fine-tuning. From Table~\ref{tab-3}, we observe that baseline CNN classifiers achieve moderate performance, with Mamba-B attaining the best results among them (F1: \textbf{82.63\%}, EER: \textbf{17.97\%} on ID/OOD). However, Table~\ref{tab-4} shows that our hybrid \textbf{\texttt{SIGNAL}} architecture significantly enhances performance across all PTMs. Mamba-B again leads with the highest F1-score of \textbf{90.99\%} and the lowest EER of \textbf{7.92\%} under the ID/OOD split, demonstrating strong generalization to out-of-domain singing data. Even smaller models such as Mamba-S benefit considerably, reducing EER to \textbf{10.38\%} with the hybrid setup. These findings affirm the robustness and transferability of \textbf{\texttt{SIGNAL}} to domains with different acoustic structures, such as musical and multilingual content.
\begin{figure*}[!hbt]
    \centering
    \includegraphics[width=0.78\linewidth]{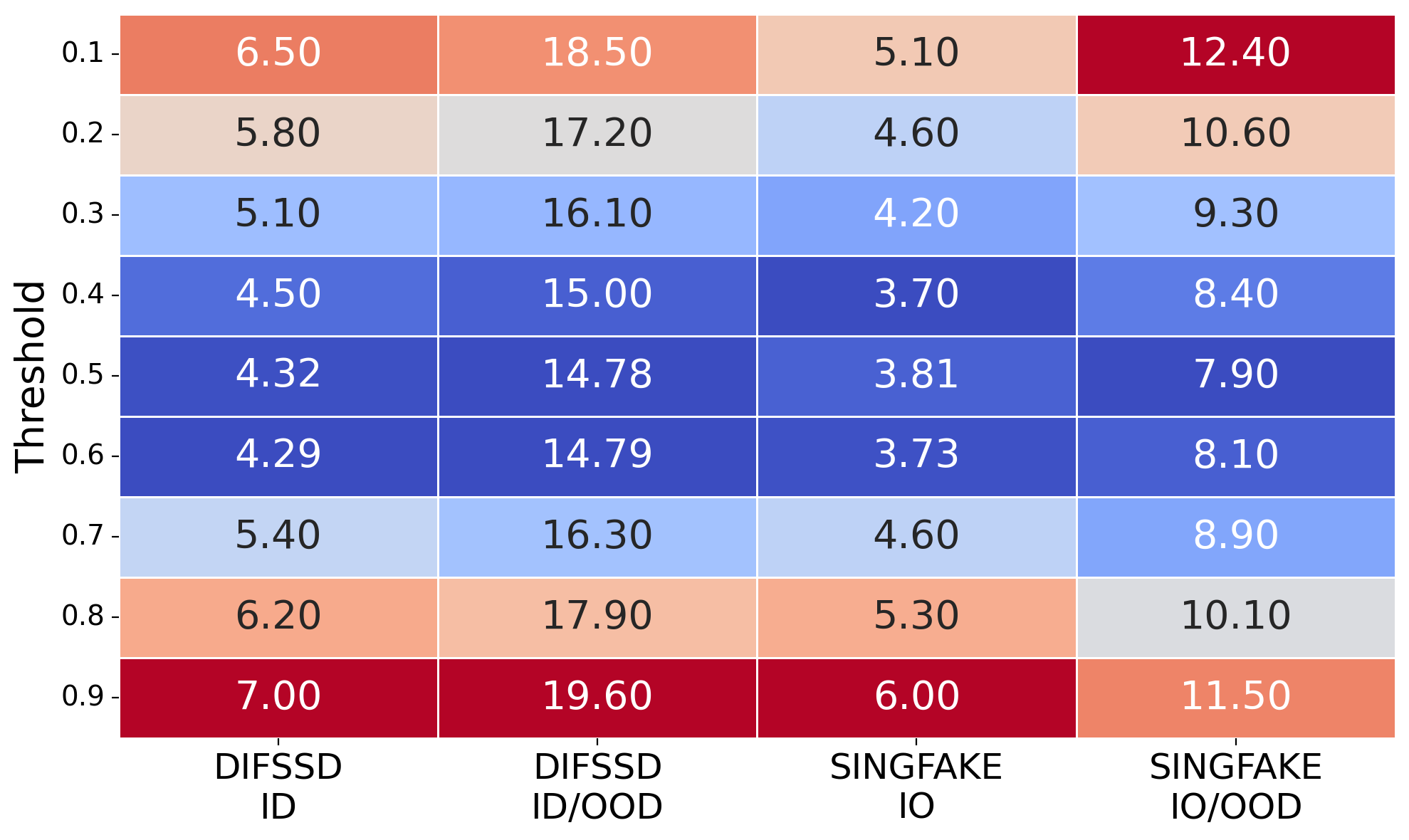}
    % \caption{MAMBA-B GNN+KNN diffssd/Singfake results on threshold}
    \caption{Threshold $\tau$ sensitivity of \textbf{\texttt{SIGNAL}} (KNN+GNN) with Mamba-B.}
    \label{fig:enter-label}  
\end{figure*}
\subsection{Ablation Study}
\label{ablatinghgho}
To evaluate the impact of threshold selection in our GNN+KNN hybrid architecture, we perform an ablation analysis on both the DiffSSD and SingFake datasets using the best-performing representation, Mamba-B. As described earlier, our model applies a confidence-based routing mechanism, where samples with prediction confidence above a threshold $\tau$ are classified as originating from seen generators, while those below $\tau$ are flagged as unseen. Figure~\ref{fig:enter-label} presents the Equal Error Rate (EER) across a range of threshold values ($\tau \in [0.1, 0.9]$), evaluated on four evaluation splits: DiffSSD ID, DiffSSD ID/OOD, SingFake ID, and SingFake ID/OOD. The results clearly indicate that $\tau=0.5$ offers a balanced trade-off—minimizing EER on both in-distribution and open-set splits. This analysis validates our design choice of setting $\tau=0.5$ as the open-set decision boundary in \textbf{\texttt{SIGNAL}}.

\section{Conclusion}
% \vspace{-0.458cm}
This work address the dual challenge of synthetic speech attribution and detection of unseen generators. We show that Mamba-based representations are highly effective in capturing generator-specific traits, including prosodic patterns and synthesis artifacts, owing to their advanced temporal modeling. Building on this insight, we introduce a hybrid framework \textbf{\texttt{SIGNAL}} that combines graph-based relational modeling with instance-level KNN inference to exploit both global structure and local similarity. Our study highlights the overlooked potential of graph-enhanced modeling for robust synthetic speech detection, providing a strong foundation for advancing attribution and generalization in generative speech forensics. This work sets a solid baseline for future research in structured and generalizable synthetic speech analysis. \newline
\section{Limitations and Future Work}  
First, our evaluation is conducted on two publicly available benchmarks (DiffSSD and SingFake), which, while diverse, do not cover all possible synthesis conditions or generators.
Second, SIGNAL relies on a decision threshold for open-set detection; although empirically stable, threshold selection remains a challenge in fully unconstrained deployment scenarios.
Third, our current framework identifies whether a sample originates from an unseen generator but does not perform fine-grained attribution among multiple unseen sources. Extending SIGNAL toward hierarchical or cluster-based modeling of unseen generators is a promising direction for future work. \newline
\noindent \textbf{Interpretability and Explainability:} Although the framework achieves strong performance, its decisions remain somewhat opaque. The current framework lacks mechanisms for explaining why a particular speech sample is attributed to a specific generator or flagged as unseen. This limits its usability in high-stakes scenarios where transparency is essential. Enhancing interpretability—through attention visualization or prototype-level explanations—is an important direction for future work. \newline
\noindent \textbf{Threshold sensitivity:}
SIGNAL relies on a decision threshold to balance attribution and open-set detection. While Figure~\ref{fig:enter-label} shows stable performance across a wide range of values, we acknowledge that no fixed threshold is universally optimal under all real-world deployment conditions. Adaptive or data-driven threshold selection remains an important direction for future work.
\section{Ethical Statement}
This study addresses the growing concern around synthetic speech and singing voice generation by proposing methods to detect and attribute artificially generated content. We use only publicly available datasets (DiffSSD and SingFake), and no personal or sensitive data is involved. Our framework is developed purely for detection and research purposes—not for generating or misusing synthetic content.

\bibliography{main}

\appendix

% \section{Appendix}
% \label{sec:appendix}

% \subsection*{Implementation Details and Hyperparameters}

% \paragraph{CNN projection head ($f_{\text{cnn}}$):} We project pooled SFM embeddings to $d=64$ with a 1D CNN encoder:
% Conv1 (channels 256, kernel 3, stride 1) $\rightarrow$ ReLU $\rightarrow$ MaxPool (2),
% Conv2 (channels 128, kernel 3, stride 1) $\rightarrow$ ReLU $\rightarrow$ MaxPool (2),
% followed by a dense layer to 64 and dropout 0.1.

% \paragraph{GNN head:} We use multi-head self-attention over class prototype nodes with 4 heads, hidden size 64, attention dropout 0.1, and residual connections. Logits are computed via a linear projection to $N$ generator classes. Attention entropy $\mathcal{H}_{\text{attn}}$ is computed from $\mathbf{p}_{\text{GNN}}$.

% \paragraph{KNN branch:} We use distance-weighted KNN with $K=5$ neighbors and squared $\ell_2$ distance. We set $\epsilon=10^{-8}$ for numerical stability in inverse-distance weighting.

% \paragraph{Fusion and routing:} We fuse posteriors with $\alpha=0.5$ and use confidence routing with default threshold $\tau=0.5$. We do not use an entropy threshold for model selection in the main experiments.

% \paragraph{Optimization:} We use Adam (lr $1\times10^{-3}$, weight decay $1\times10^{-5}$), batch size 32, and train for up to 50 epochs with early stopping on dev EER (patience 5). We set the random seed to 42 and report results from a single run per setting.

\end{document}